\documentclass[12pt]{article}
\usepackage{epsf}
\usepackage{graphicx}
\usepackage{dcolumn}
\usepackage{bbm}
\newcommand{\bi}{\bigskip}
\newcommand{\no}{\noindent}
\newcommand{\be}{\begin{eqnarray}}
\newcommand{\ee}{\end{eqnarray}}
\newcommand{\hk}{\hspace{0.1cm}}

\newcommand{\rk}{\right)}
\newcommand{\lk}{\left(}

\newcommand{\il}{\int\limits}

\begin{document}

\title{On the Yang-Mills wave functional in Coulomb gauge}
\date{\today}

\author{H. Reinhardt and C. Feuchter\\
Institut f\"ur Theoretische Physik\\
Auf der Morgenstelle 14\\
D-72076 T\"ubingen\\
Germany}
%



\bi

\no
\maketitle
\bi

\begin{abstract}
We investigate the dependence of the Yang-Mills wave functional in Coulomb gauge
on the Faddeev-Popov determinant. We use a Gaussian wave functional multiplied
by an arbitrary power of the Faddeev-Popov determinant. We show, that within the
resummation of one-loop diagrams the stationary vacuum energy is independent of
the power of the Faddeev-Popov determinant and, furthermore, the wave
functional becomes field-independent in the infrared, describing a stochastic vacuum. 
Our investigations show,
that the infrared limit is rather robust against details of the variational
ans\"atze for the Yang-Mills wave
functional. The infrared limit is exclusively 
determined by the divergence of the 
Faddeev-Popov
determinant at the Gribov horizon.
\end{abstract}

\noindent
{\bf pacs:} {11.10Ef, 12.38Aw, 12.38 Cy, 12.38 Lg}

\no
\section{Introduction}
%
Recently Yang-Mills theory in Coulomb gauge has become the subject of intensive
studies both on the lattice, refs. \cite{1}, \cite{2} and in the continuum, ref.
\cite{3}, \cite{4}, \cite{5}, \cite{6}, \cite{7}. The Coulomb gauge is a
physical gauge and in this gauge confinement is realized by the statistical
dominance of the field configurations near the Gribov horizon, which gives rise
to an infrared enhanced static color charge potential. 

For the calculation of static properties of continuum Yang-Mills theory, the
Schr\"odinger equation approach \cite{RX2} seems to be most convenient. In refs. \cite{5},
\cite{6}, \cite{7} the Yang-Mills Schr\"odinger equation was approximately solved
in Coulomb gauge, using the variational principle. Different ans\"atze for the
vacuum wave functional and different renormalization conditions have been used, and different infrared 
behaviours of the gluon and
ghost propagators were obtained. One might wonder, whether the different results
are a consequence of the different ans\"atzes for the wave functional. To answer
this question, in this paper we consider a more general class of wave
functionals, which includes, in particular, the wave functionals previously used
in refs. \cite{5},
\cite{6} and \cite{7}. We will show, that the different ans\"atze used so far, 
have to yield
the same unique infrared behaviour of the vacuum wave functional (at least to
the order considered). We will also show, that in the infrared the wave
functional becomes field independent, describing a stochastic vacuum.
Furthermore, the infrared limit of the wave functional agrees with
 the exact vacuum wave functional in 
$D= 1 + 1$.
\bi

\no
\section{The variational ansatz}
\bi

\no
For the Yang-Mills vacuum we consider trial wave functionals of the form
\be
\label{1}
\Psi [A^\perp] = J^{- \alpha} [A^\perp] \phi [A^\perp] \hk ,
\ee
where 
\be
\label{2}
J [A^\perp]= \frac{Det(- \hat{D}_i \partial_i)}{Det(- \partial^2 )}
\ee
is the Faddeev-Popov determinant, which for later convenience has been
normalized to $J [A^\perp=0]=1$. Here $ \hat{D} = \partial + \hat{A}^\perp$ is the covariant
derivative and $\hat{A}^\perp = A^{\perp a} \hat{T}^a$ denotes the gauge field in the adjoint
representation.
Furthermore $\phi [A^\perp]$ is a Gaussian wave functional defined by
\be
\label{3}
\phi [A^\perp] & = & {\cal N} \exp (- S [A^\perp]) \nonumber\\
S[A^\perp] & = & \frac{1}{2} \int d^{3}x \int d^{3}x' A^{\perp a}_{i} ({\bf x}) 
\omega^{ab}_{ij} ({\bf x},{\bf x}') A^{\perp b}_{j} ({\bf x}')
\ee
with ${\cal N}={\cal N}( \alpha , \omega)$ being 
a normalization constant to ensure 
$\langle \psi \mid \psi \rangle = 1 $. 
The Faddeev-Popov determinant arises as Jacobian in
the transformation from ``cartesian'' 
coordinates $A^{a}_{i} ({\bf x})$ to the ``curvilinear
coordinates'' 
$A^{\perp a}_{i} ({\bf x})$ satisfying the Coulomb gauge $\partial_i
A^\perp_i = 0$, and defines the metric in
the space of transversal gauge orbits $A^{\perp}_{i} ({\bf x})$. Accordingly the
scalar product in the space of transversal gauge orbits is defined by
\be
\label{4}
\langle \Psi \mid \Phi \rangle = \int\limits D A^{\perp} J [A^{\perp}]
\Psi^* [A^\perp] \Phi [A^\perp] \hk ,
\ee
where the integration should in principal be restricted to the fundamental
modular region \cite{4}, \cite{13}. The choice $\omega^{a b}_{i j} ({\bf x}, {\bf x}') = \delta_{i j}
\delta^{a b} \omega ({\bf x}, {\bf x}')$ and $\alpha = 0$ was used in refs. \cite{5},
\cite{6}, while
$\alpha = \frac{1}{2}$ was chosen in ref. \cite{7}. From eq. (\ref{1}) it is
seen, that for the latter
 choice $\phi [A^\perp]$ represents just the ``radial'' wave functional\footnote{For a point particle in a $s$-state the wave function is of the
 form $\Psi (r) = \frac{\phi (r)}{r}$, where $\phi (r)$ is the radial wave function and the Jacobian is given by $J = r^2$}. 

We wish to study the dependence of the vacuum
Yang-Mills wave functional (1) on the power of the Faddeev-Popov determinant
$\alpha$, which can take, in principle,
 any real value as long as $\Psi [A^\perp]$ is normalizable.
The integral kernel $\omega$ as well as
the parameter $\alpha$ have to be determined by minimizing the expectation value
of the energy
\be
\label{5}
\langle H \rangle = \int D A^\perp J [A^\perp] \Psi^* [A^\perp] H \Psi [A^\perp] \hk .
\ee
\bi

\no
\section{Minimization of the energy}
\bi

\no
Inserting the explicit form of the wave functional eq. (\ref{1}) into eq.
(\ref{5}) variation of the energy with respect to the kernel $\omega$ yields 
the ``gap equation'' :
\be
\label{6}
\frac{\delta \langle H \rangle}{\delta \omega}= 2 \left \langle \frac{\delta S}{\delta
\omega} \right \rangle \left \langle H \right \rangle - \left \langle \left\{ \frac{\delta S}{\delta 
\omega}, H \right\}
\right \rangle = 0 \hk  .
\ee
Here the first term arises from the variation of the normalization constant
$(\delta {\cal N} / \delta \omega = {\cal N} \langle \delta S / \delta \omega
\rangle)$ and $\{ , \}$ denotes the anti-commutator.
Minimization of the energy $\langle H \rangle$ (\ref{5})
with respect to the power $\alpha$ yields the
condition 
\be
\label{7}
\frac{d \langle H \rangle}{d \alpha} = 2 \langle \ln J[A^\perp] \rangle
\langle H \rangle - \left \langle \left\{ \ln J[A^\perp],H \right\} \right \rangle = 0 \hk ,
\ee
where we have used $d {\cal N} / d \alpha = {\cal N} \langle \ln J[A^\perp] \rangle$. \\
%
Consider now the structure of
the Faddeev-Popov determinant (2), which obviously satisfies $\ln J[A^\perp=0] = 0$.
Furthermore, by definition (\ref{2})
we have 
\be
\label{5a}
\frac{\delta \ln J[A^\perp]}{\delta A^{\perp a}_{i} ({\bf x})} = - Tr \left( G \Gamma^{o,a}_{k} ({\bf x}) \right) \hk ,
  \ee
where
\be
\label{8} 
G = (-\hat{D}_i \partial_i)^{-1}
\ee
is the inverse Faddeev-Popov operator and $\Gamma^{0,a}_{k} ({\bf x})= \delta G^{-1}
/ \delta A^{\perp a}_k ({\bf x})$ is the bare
ghost gluon vertex  \cite{7}. 
Since  $\Gamma^{0,a} \sim \hat{T}^{a}$ (group generator
in the adjoint representation) and $\hat{T}^{a}$ occurs in $G$ only in the
combination $\hat{A}^\perp = A^{\perp a} \hat{T}_{a}$ it is clear, that the quantity
(\ref{5a})
has to be proportional to $A^\perp$, since $tr \hat{T}^{a} = 0$. Therefore we find the
representation
\be
\label{9}
\ln J[A^\perp] = \int d^{3}x d^{3}x' C^{ab}_{ij} [A^\perp]({\bf x},{\bf x}') A^{\perp a}_{i} 
({\bf x}) A^{\perp b}_{j} ({\bf x}')\hk \hk  \hk 
\ee
with some, not explicitly known functional $C^{ab}_{ij} [A^\perp]$. In one-loop
approximation the expectation value of eq. (\ref{9}) is given by
\be
\label{10}
\langle \ln J[A^\perp] \rangle \simeq \int  d^{3}x d^{3}x' \left \langle C^{ab}_{ij}
[A^\perp]({\bf x},{\bf x}') \right \rangle \nonumber\\
\cdot \left \langle A^{\perp a}_{i} ({\bf x}) A^{\perp b}_{j} ({\bf x}') \right \rangle \hk .
\ee
Furthermore, to this order we can neglect terms of the form $ \left \langle \frac{\delta C}{\delta A^\perp} A^\perp \right \rangle$ 
and $\left \langle \frac{\delta^{2} C}{\delta A^\perp \delta A^\perp }
\right \rangle$ and find from (\ref{9}) 
for the
curvature in orbit space \cite{7}
\be
\label{15}
\chi^{ab}_{ik}({\bf x},{\bf x}') &=& - \frac{1}{2} \left \langle \frac{\delta^{2} \ln J}{\delta A^{\perp a}_{i}
({\bf x})  \delta A^{\perp b}_{j} ({\bf x}')} \right \rangle \nonumber\\
&=& - \left \langle C^{ab}_{ij} [A^\perp] ({\bf x},{\bf x}') \right 
\rangle \hk .
\ee 
To this order we can also replace $C [A]$ in (\ref{10}) by its expectation value
$(- \chi)$ yielding
\be
\label{XX}
\ln J [A^\perp] = -  \int d^3 x d^3 x' \chi^{a b}_{i j} ({\bf x}, {\bf x}') 
A^{\perp a}_i ({\bf x}) A^{\perp b}_j ({\bf x}') \hk .
\ee
Inserting eq. (\ref{XX}) into eq. (\ref{7}) we find
\be
\label{11}
\frac{d \langle H \rangle}{d \alpha} = - \int  d^{3}x d^{3}x'  \chi^{ab}_{ij}
({\bf x},{\bf x}') \Big[ 2 \left \langle  A^{\perp a}_{i} ({\bf x}) A^{\perp b}_{j} ({\bf x}') \right \rangle \left \langle H
\right \rangle - \left \langle \left \{ A^{\perp a}_{i} ({\bf x}) A^{\perp b}_{j} ({\bf x}'), H \right \} \right \rangle \Big] \hk \hk \hk
\ee
On the other hand for the Gaussian wave functional (\ref{1}), (\ref{3}) we have 
\be
\label{12}
\frac{\delta S}{\delta \omega^{ab}_{ij}({\bf x},{\bf x}')} = \frac{1}{2} A^{\perp a}_{i} ({\bf x}) A^{\perp b}_{j}
({\bf x}') \hk ,
\ee
so that the equation (\ref{6}) becomes
\be
\label{13}
2 \frac{\delta \langle H \rangle}{\delta \omega^{ab}_{ij}({\bf x},{\bf x}')} = 2 \left \langle
A^{\perp a}_{i} ({\bf x}) A^{\perp b}_{j} ({\bf x}') \right \rangle \left \langle H \right \rangle - \left \langle \left\{ 
A^{\perp a}_{i} ({\bf x}) A^{\perp b}_{j} ({\bf x}') H \right\} \right \rangle \hk .
\ee
Comparison of eqs. (\ref{11}) and (\ref{13}) yields
\be
\label{14}
\frac{d \langle H \rangle}{d \alpha} = - 2 \int d^{3}x d^{3}x' \chi^{ab}_{ij}
({\bf x},{\bf x}')  \frac{\delta \langle H \rangle}{\delta
\omega^{ab}_{ij}({\bf x},{\bf x}')} \hk .
\ee
Thus stationarity of the energy with respect to $\omega^{ab}_{ij}({\bf x}) \hk , \hk
 \delta \langle H \rangle / \delta \omega =0$ 
implies also stationarity with respect to $\alpha, d
\langle H \rangle / d \alpha = 0$. Let us emphasize, that eq. (\ref{14}) 
is exact to one-loop
order in the equation of motion (i.e. to two-loop order in $\langle H \rangle$).
\bi

\no
\section{The energy functional}
\bi

\no 
The above obtained result can be also immediately infered 
from the explicit
expression of the expectation value of the Yang-Mills Hamiltonian in the state
(\ref{1}), which is given by
\be
\label{17}
\langle H \rangle &  = & E_k + E_B + E_C \\
\nonumber\\
E_k & = & \delta^3 (0) \frac{N^2_C - 1}{2} 
          \int d^3 k \frac{\left[ \Omega ({\bf k}) 
	  - \chi ({\bf k}) \right]^2}{\Omega ({\bf k})} 
	  \nonumber\\
E_B & = & \delta^3 (0) \frac{N^2_C - 1}{2} 
          \int d^3 k  \left( \frac{{\bf k}^2}{\Omega
	  ({\bf k})} + \frac {N_C g^2} {8} \int \frac{d^3 k'}{(2 \pi)^3} 
	  \left[ 3 - ( \hat{{\bf k}} \hat{{\bf k}'} )^2 \right] 
	  \frac {1} {\Omega ({\bf k}) \Omega ({\bf k}')} \right) 
	  \nonumber\\
E_C & = & \delta^3 (0) \frac{N_C (N_C^2 - 1)}{16} 
          \int \frac{d^3 k d^3 k'}{(2 \pi)^3} 
	  \left[ 1 + ( \hat{{\bf k}} \hat{{\bf k}'} )^2 \right] 
	  \cdot 
	  \frac{d^2 ({\bf k} + {\bf k}') f ({\bf k} + {\bf k}') 
	  }{({\bf k} + {\bf k}')^2} 
	  \cdot 
	  \nonumber\\
&& \phantom{\delta^3 (0) \frac{N_C (N_C^2 - 1)}{16} 
          \int \frac{d^3 k d^3 k'}{(2 \pi)^3}} 
          \cdot	   
	  \frac{\left[ \left( \Omega ({\bf k}) - \chi ({\bf k}) \right) 
	  - \left(\Omega ({\bf k}') - \chi ({\bf k}') \right)
	  \right]^2}{\Omega ({\bf k}) \Omega ({\bf k}')} \hk , 
	  \nonumber
\ee
where $d ({\bf k})$ and $f ({\bf k})$ are the ghost and Coulomb form factors defined in ref.
\cite{7} and $\chi$ is the scalar curvature defined in terms of the curvature
tensor (\ref{15}) by
\be
t_{k n} (x) \chi^{a b}_{n l} ({\bf x}, {\bf y}) = \delta^{a b} t_{k l} ({\bf x}) \chi ({\bf x}, {\bf y}) 
\ee
\be
\label{F6}
\chi ({\bf k}) = \frac{N_C}{4} \int \frac{d^3 q}{(2 \pi)^3} \left[ 1 - ( \hat{{\bf k}} \hat{{\bf q}} )^2
\right] \frac{d ({\bf k} - {\bf q}) d ({\bf q})}{ \left({\bf k}- {\bf q} \right)^2}
\ee
with $t_{k l} ({\bf x}) = \delta_{k l} - \partial_k \partial_l / \partial^2$ being the
transversal projector
\footnote{The ghost and Coulomb form factor, $d(k)$ and $f(k)$, satisfy the Schwinger-Dyson
equation derived in ref. \cite{7} with $\omega$ replaced by $\Omega$.}.
Furthermore
\be
\label{21a}
\Omega ({\bf k}) = \omega ({\bf k}) - (2 \alpha - 1) \chi ({\bf k})
\ee
is the inverse of the gluon propagator
\be
\label{5XX}
\langle A^{\perp a}_i ({\bf k}) A^{\perp b}_j (- {\bf k}) \rangle = \frac{1}{2} \delta^{a b} t_{i j} ({\bf k})
\Omega^{- 1} ({\bf k}) \hk .
\ee
Note, that the curvature $\chi ({\bf k})$ (\ref{F6}) is entirely determined by the
ghost form factor $d ({\bf k})$ and does not depend on $\omega ({\bf k})$. 
The energy (\ref{17}) 
depends on $\alpha$ and $\omega ({\bf k})$ only through the combination $\Omega ({\bf k}) = \omega ({\bf k})
- (2 \alpha - 1) \chi ({\bf k})$. From this fact immediately follows, that eq.
(\ref{6}) implies eq. (\ref{7}), so we find again, that the wave
functional (\ref{1}) which minimizes the energy is independent of $\alpha$. In
fact, since $\langle H \rangle$ (\ref{17}) depends on $\omega$ and $\alpha$ only
through the combination $\Omega = \omega - (2 \alpha - 1) \chi$ it suffice to
minimize the energy with respect to $\Omega$. The resulting gap equation
\footnote{Although the curvature (\ref{F6}) does not explicitly
depend on $\omega$ it depends implicitly on $\omega$ via the ghost form factor 
$d(k)$. However, also the ghost form factor $d (k)$ depends on $\omega$ only
through $\Omega$. This dependence is, 
however, a higher
order effect and to one-loop order $\frac{\delta \chi}{\delta \omega}$ or
$\frac{\delta \chi}{\delta \Omega}$ can be neglected.
Then the gap equation $\frac{\delta \langle H \rangle }{\delta \Omega} = 0$ is exactly
the one obtained in ref. \cite{7} with $\omega$ replaced by $\Omega$ .}
also depends only on $\Omega$ and its solution  is
independent of $\alpha$.
This shows, that the infrared behaviour of the gluon propagator 
$\langle A^\perp
A^\perp \rangle $ (\ref{5XX}) is independent of the power $\alpha$
 of the Faddeev-Popov determinant assumed in the wave
functional (\ref{1}). 
Therefore we are free to choose $\alpha$ for
our convenience, for example, $\alpha = \frac{1}{2}$.  This choice has the
technical advantage, that $\Omega ({\bf k}) = \omega({\bf k})$, which allows a
straightforward application of Wick's theorem in the calculation of expectation
values. \\
%
In this context let us also mention, that the choice $\alpha =
\frac{1}{2}$ in eq. (\ref{1}) yields the wave function used by the present
authors in ref. \cite{7}, while the wave function used in refs. \cite{5},
\cite{6}
corresponds to the choice $\alpha = 0$. Inspite of the different wave functions
chosen in refs.  \cite{5},
\cite{6} and ref. \cite{7} the same infrared behaviour of the gluon
propagator should be obtained  in one-loop order, as shown above, provided the same renormalization 
condition is used. 
However, while refs. \cite{5},
\cite{6} finds an
infrared finite gluon propagator, we find an infrared vanishing gluon
propagator \cite{7}. 
Two sources of the different behaviours obtained in 
ref. \cite{7}, and refs. \cite{5}, \cite{6} 
come to mind:
i) different choices of the renormalization condition and
ii) different treatments of the curvature of orbit space.
In ref. \cite{7} we choose the socalled horizon condition 
\cite{13}
\be
\label{30}
d^{-1}(k \to 0)=0
\ee
as renormalization condition while refs. \cite{5}, \cite{6} require the kernel $\omega$
in the Gaussian wave functional to be infrared finite $\omega (k \to 0) = \textrm{const.}$
\footnote{In $D=2+1$ we find
a self-consistent solution to the coupled Schwinger-Dyson equations only when we impose
the horizon condition (\ref{30}).}.
Furthermore, while the curvature of orbit space $\chi$ (\ref{15}) was fully 
included in
ref. \cite{7}, it was completely neglected in \cite{5} and ignored \footnote{In
the formal part of \cite{6} the curvature was fully included.}  
in the Coulomb energy
in the numerical calculations of ref. \cite{6}.
As will be shown in the next section ignoring the curvature in the Coulomb energy will 
change the infrared behaviour of the wave functional.
It was already observed in ref. \cite{7}, that
the full inclusion of the curvature is vital for the infrared limit of the
theory. This is consistent with the observation in Landau gauge, that in the
Schwinger-Dyson equations the ghost
loop is by far more important than the gluon loop \cite{RX1}.
\bi

\no
\section{The Yang-Mills wave functional in the infrared}
\bi

\no
With the relation  (\ref{XX}) 
the wave functional (\ref{1}) becomes
\be
\label{21}
\Psi [A^\perp] \simeq e^{\alpha \int A^\perp \chi A^\perp - \frac{1}{2} \int A^\perp \omega A^\perp} \hk .
\ee
Furthermore, the solution of the gap equation eq. (\ref{6}) is such, that in 
the
infrared
\be
\label{22}
\chi (k \to 0) = \Omega (k \to 0) \hspace{0.2cm} , \hspace{0.2cm} 
2 \alpha \chi (k \to 0) = \omega (k \to 0)
\ee
holds. This is an extension of the relation $\chi (k \to 0) = \omega (k \to 0)$
found in ref. \cite{7} for $\alpha = \frac{1}{2}$.
 With the relation (\ref{22})  the vacuum Yang-Mills 
wave functional becomes in the infrared  
\be
\label{23}
\Psi [A^\perp] = 1
\hk .
\ee
In ref. \cite{9} this wave functional was assumed in the infrared 
regime, for sake of simplicity.
We have thus shown, that, to one-loop order, eq. (\ref{23}) is the correct wave
function in the infrared. \\
%
The infrared wave functional $\Psi [A^\perp] = 1$ means, that gauge
fields at distant positions ${\bf x}, {\bf x}' , | {\bf x} - {\bf x}' | \to
\infty$ are completely uncorrelated. Accordingly, the gluon propagator $\langle
A^\perp ({\bf x}) A^\perp ({\bf x}') \rangle$ has to vanish rapidly in the infrared $| {\bf x} - {\bf x}' | \to
\infty$, which is in agreement with the findings of ref. \cite{7}. Thus, the wave functional $\Psi [A^\perp] = 1$ 
describes a  stochastic
vacuum, in which color (correlation) cannot propagate over large distances. This is
nothing but color confinement. In this sense, the infrared wave functional
(\ref{23}) supports the picture of a stochastic Yang-Mills vacuum \cite{10}. \\
%
Let us also stress, that in view of the relation (\ref{22}) the infrared
behaviour of the gluon propagator (\ref{5XX}) is exclusively determined by the
curvature $\chi$ (\ref{15}) in orbit space. Furthermore from eqs. (\ref{23}) and
(\ref{XX}) follows that in the infrared limit
the vacuum expectation values are described by an Gaussian ensemble
\be
\langle \cdots \rangle = \int {\cal D} A^\perp J [A^\perp] \cdots = \int D A^\perp \cdots e^{- \int
A^\perp \chi A^\perp} \hk . \hk \hk \hk
\ee
Finally, let us also note that the relation (\ref{22}) holds independent of the employed 
renormalization condition as long as the curvature $\chi (k)$ is infrared divergent.
Since the Faddeev-Popov determinant vanishes on the Gribov horizon, which contains the infrared
dominant field configurations, from eq. (\ref{XX}) follows that the curvature has, indeed, to be infrared divergent.
The condition (\ref{22}) is, however, lost when the curvature is neglected in the Coulomb energy as done in ref. 
\cite{5}, \cite{6}. Given the infrared singular behaviour of $\chi (k \to 0)$ the condition (\ref{22}) implies
that for $\alpha \ne 0$ (in particular for $\alpha = \frac {1}{2}$ \cite{7}) the variational kernal $\omega (k)$
in the Gaussian ansatz (\ref{3}) has to be infrared singular, while the choice $\alpha = 0$ \cite{5}, \cite{6}
can tolerate an infrared finite $\omega (k)$.
\bi

\no
\section{Yang-Mills theory in $D = 1 + 1$}
\bi

\no
Let us test the above result in $1 + 1$ dimension, where Yang-Mills theory can
be solved exactly on a torus and reduces to quantum mechanics in  curved space. \\
%
Implementing the Coulomb gauge $\partial_1 A_1 = 0$, there is only a constant
gauge field $A_1 (x_1) = \textrm{const.}$ left, which can be diagonalized in color space
 by
exploiting the residual global gauge freedom $U$, 
not fixed by $\partial_1 A_1 = 0$.
Defining the remaining quantum mechanical degree of freedom, $a$, by
\be
g A_1 L \equiv g L A^a_1 \frac{\tau^a}{2}  = U
\frac{a}{2} \tau^3 U^\dagger \hk ,
\ee
where $L$ is spatial extension of the torus
the Faddeev-Popov determinant becomes \cite{11}
\be
\label{105}
J (a) = \sin^2 a \hk .
\ee
The Gribov horizon occurs at $a = n \pi$ and the fundamental modular region is
obviously given by $0 \leq a \leq \pi$. Furthermore the Yang-Mills Hamiltonian in the
variable $a$ is given by
\be
\label{XXa}
H_{kin} = - \frac{g^2 L}{8} \frac{1}{\sin^2 a} \frac{d}{d a} \sin^2 a \frac{d}{d
a} \hk .
\ee
In one spatial dimension there is no magnetic field and no dynamical gluon 
charge $(- \hat{A}^{a b}_1 \Pi^a_1 = 0)$, since the gauge field has only one
(non-zero) color degree of freedom.
Accordingly, the Coulomb term of the Yang-Mills Hamiltonian \cite{RX2} 
vanishes in the
absence of external color charges. \\
%
With the ansatz
\be
\label{Z1}
\Psi_k (a) = \frac{1}{\sqrt{J (a)}} \phi_k (a) = \frac{1}{\sin a} \phi_k (a) \hk ,
\ee
which corresponds to the choice $\alpha = \frac{1}{2}$ in eq. (\ref{1}), 
the Schr\"odinger equation
$H \Psi_k = E_k \Psi_k$
reduces to
\be
- \frac{g^2 L}{8} \phi''_k (a) = \lk E_k + \frac{g^2 L}{8} \rk \phi_k (a) \hk ,
\ee
whose solution is given by\footnote{This solution was previously found in ref.
\cite{12}}
\be
\label{ZZ}
\phi_k (a) = \sin (k a) \hk , \hk E_k = \frac{g^2 L}{8} (k^2 - 1) \hk .
\ee
In the continuum limit $L \to \infty $ only the vacuum state $k = 1$ survives $(E_1 =
0)$, while all excited states $k > 1$ aquire an infinite energy and are thus
frozen out. The vacuum wave function is given by (\ref{Z1}), (\ref{ZZ})
\be
\Psi_{k = 1} (a) = 1 \hk ,
\ee
which is precisley the infrared limit of the vacuum Yang-Mills wave functional
in $D = 3 + 1$ found above (see eq. (\ref{23})). 
Note also that the radial wave function $\phi_k (a)$ (\ref{ZZ}) vanishes on the Gribov horizon $a = n \pi $
to compensate for the vanishing of the Faddeev-Popov determinant $J (a)$ (\ref{105}), just like in the $D=3+1$
dimensional case where (for $\alpha = \frac {1}{2}$), the radial wave functional $\phi [A^{\perp}]$
(\ref{3}) vanishes in the infrared due to the infrared divergence of $\omega (k \to 0)$.

\bi

\no
\section{Summary and Conclusions}
\bi

\no
We have studied the variational solution of the Yang-Mills Sch\"odinger equation
in Coulomb gauge for a class of wave functionals (\ref{1}) 
consisting of a Gaussian and an arbitrary power ($- \alpha$) of the
Faddeev-Popov determinant. We have found, that up to one-loop in the gap equation
(i.e. two loops in the energy) the stationary solution is independent of this
power $\alpha$. The same is true for the transversal gluon propagator 
(\ref{5XX}) which is exclusively determined by the self-consistent solution $\Omega$ of the 
gap equation $\frac{\delta \langle H \rangle}{\delta \omega} = 0$. This solution $\Omega$ 
is independent of the choice of $\alpha$. Different choices of $\alpha$ will lead to different kernels 
$\omega$ (with possibly different infrared behaviours) in the wave functional
 (\ref{3}). 
But this will not affect the gluon propagator (\ref{5XX}).
Furthermore in the infrared the Yang-Mills vacuum wave-functional becomes
field-independent describing a stochastic vacuum, in which color cannot
propagate over large distances. The infrared limit of the wave functional becomes
exact in $D = 1 + 1$. \\
Our investigations show, that the infrared behaviour of
Yang-Mills theory in Coulomb gauge is rather robust with respect to changes in
the variational ans\"atze for the wave functional as long as the curvature in
orbit space induced by the Faddeev-Popov determinant is properly included.
\bi

\no
\section*{Acknowledgements}
\bi

\no
The authors are grateful to R. Alkofer, J. Pawlowski, D. Zwanziger and, in particular, to A. Szczepaniak for
a critical reading of the manuscript and useful comments. One of the authors (H.R.) also acknowledges
stimulating discussions with A. Szczepaniak during the ``QCD Down under workshop'' at the CSSM in Adelaide,
where this work was started. He thanks the CSSM for the warm hospitality extended to him and for financial
support.
This work was partially supported by Deutsche Forschungs-Gemeinschaft under contract DFG-Re 
856.

\no

\end{document}